\documentstyle[seceq,preprint,epsf]{ptptex}

\title{
Quest for the Quark-Gluon Plasma\footnote{Invited talk given at Tokyo-Adelaide Joint
Workshop on Quarks, Astrophysics and Space Physics, University of Tokyo, Jan. 6 -10, 2003.}
}

\author{
T.~Matsui\footnote{E-mail address: tmatsui@nt1.c.u-tokyo.ac.jp}
}

\inst{
Institute of Physics, University of Tokyo \\
Komaba, Tokyo 153-8902, Japan}

\recdate{
\today
}

\abst{
Present status and future prospect of the quest for the quark-gluon plasma, 
the primordial form of matter which once pervaded the early universe, 
with ultrarelativistic nuclear collisions are discussed.
}

\begin{document}

\maketitle

\section{Introduction}
With the advent of the Relativistic Heavy-Ion Collider (RHIC) at Brookhaven 
we are now able to study nuclear collisions at extreme relativistic energies of up to 200$A$ GeV
with heavy ions as heavy as gold nuclei ($A = 197$).  
These energies far exceed the rest mass energy of the nucleon, $m_N c^2 = 0.94$GeV,
hence the term  "ultrarelativistic" applies.
The primary physics motivation of studying nuclear collision at such ultrarelativistic energies is
to recreate the physical conditions similar to those which once prevailed in the very early universe
and study the primordial form of matter 
from which all the matter of the present universe was created.  
Indeed, such a nuclear collision typically results in production of more than few thousands of particles each 
having energies of 1 GeV; if these particles are in statistical equilibrium at a certain stage of the 
collision, then the temperature of the matter would be of the order of $10^{12}$ degree in Kelvin,
the temperature of the universe as early as ten microsecond after the beginning of the Big Bang. 
What is the nature of the primordial form of matter under such extreme condition?   
How matter evolved from such primordial state to the present form of matter as we see it around us? 
It is the answers to these fascinating questions which we try to learn from the experiments of
ultrarelativistic nuclear collisions.

In what follows, I discuss, after a brief introductory overview of our present theoretical understanding 
of the nature of matter under such extreme conditions and the space-time view of nuclear collisions at
ultrarelativistic energies, what we had expected for the signals of new physics in ultrarelativistic nuclear 
collisions and what we have actually learned so far from the past experiments with some remarks 
on the coming experiments.

\section{Extreme states of matter}
As we heat up matter around us, it experiences a series of drastic change in its state; 
from a solid, to a liquid, and then to a gas.  
These distinct states of matter are called {\it phases} of matter and the sudden change of phase 
at certain temperature is called {\it the phase transition}.   
As we heat up the matter further, all matter will be transformed eventually to a state called {\it plasma}, 
consisting of ions and electrons.   
This last transformation takes place gradually with increasing temperature by ionization of individual atom or 
molecule by collision; so it is not called the phase transition.  
Yet, the plasma, consisting of mobile charged particles, has distinct {\it collective} electromagnetic properties,
such as {\it screening} and {\it plasma oscillation};
hence it is sometimes called the {\it "fourth phase of matter"}.
Plasma also glows by emitting photons which are created by collisions of charged particles; all living creatures
on the glob, including ourselves, benefit very much from the flux of these photons emitted from Sun, a huge 
sphere of hot plasma bound together by gravity, as their primary energy source. 

Plasma is also formed when matter is compressed under high pressure; 
some of the electrons get freed from entrapments in an individual localized orbit and 
will form a {\it degenerate quantum plasma}.  
This change of state happens as a phase transition: the insulator-metal transition.  
The cold degenerate plasma do not glow itself, but still shines by reflecting light, or 
electromagnetic waves, at its surface. 

More than $99.9$\% of the mass of each atom resides in the very tiny region at the heart of the atom: 
the atomic nucleus. 
The nucleus is a droplet of a {\it Fermi liquid} consisting of nucleons, protons and neutrons, with maximum
size at $Z_{\rm Max} \simeq 114$ constrained by the Coulomb repulsion between the
protons and the saturation properties of nuclear force which holds the nucleons together.
Nuclear matter also makes a transition to a gaseous state as its temperature is raised about a few 
tens of MeV\footnote{Hereafter we use the energy scale for the temperature.  It can be translated 
to Kelvin scale by $1 {\rm MeV} = 10^{10}  {\rm K}$.} 
as can be achieved by low energy heavy ion collisions. 

As the temperature of nuclear matter is raised further light mesons are created but there will be no 
ionization of quarks, the constituents of hadrons, nor emission of gluons, the quantum of color 
gauge field which holds together the quarks to form hadrons, due to their outstanding property
referred to as {\it color confinement}. 
But with increasing temperature the density of these mesons will grow and since each meson is a composite
system having a finite spatial extension, they would percolate into a network of zones, 
filled with quarks, antiquarks, and gluons, which will eventually pervade the entire space.\cite{Bay79} 
If this naive expectation is realized, then nuclear matter should turn into a plasma of quarks and gluons
at sufficient high temperatures.

Precise natures of this transition is not known yet, including the existence of the phase transition or, 
if there is, the order of the transition, due to the difficulty of solving QCD in non-perturbative regime 
where this transition is expected.  
Currently prevailing prejudice, based on the results from Monte Carlo numerical simulations of
descritized versions of QCD on finite size lattice\cite{Kar02}, is that the transition is very rapid, 
at around $T \simeq 150$MeV, and is closely related to the restoration of the chiral symmetry of QCD 
which holds approximately for light quarks with small masses, but is broken spontaneously in the 
QCD vacuum.   What we are more sure is that the relevant energy scale for the creation of the
quark-gluon plasma is within the reach by ultrarelativistic nuclear collisions at RHIC, so we should 
seek for signals of the formation of the quark-gluon plasma.

\begin{figure}[ht]
  \epsfxsize=8cm  
  \centerline{\epsfbox{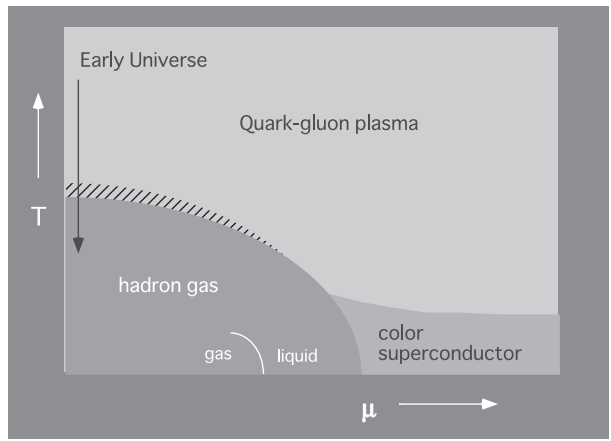}}
  \caption{Theoretical phase diagram of hot/dense matter}
   \label{fig1}
\end{figure}

Compressed form of nuclear matter may also exist in the core of compact stars, known as 
{\it neutron stars} 
which are created by the gravitational collapse of iron cores developped in massive stars, 
accompanying a spectacular event, {\it supernovae}, explosive outburst of the debris 
of the rest of the original star.   Most of the gravitational energy, of order of 10$^{53}$ erg, released 
by the formation of a neutron star,  is emitted by neutrino burst, as first confirmed in 1987 by 
Kamiokande.
The "neutron star" may be considered as a giant nuclei consisting of 
10$^{57}$ baryons (mostly neutrons) hold together by the gravitational attractive force against the 
repulsion due to the Pauli exclusion principle and nuclear force.  It is covered by a thin layer (crust) of 
a metalic form of atoms, while the central density of the neutron core may exceed ten times
the central density of ordinary nuclei.  It is thus natural to speculate that matter at the heart of 
neutron star would have melted into a dense degenerate plasma of quarks.
Such conjecture was entertained even long before the discovery of QCD.\cite{ZN67}  

It is known that in the presence of attractive two particle interaction, irrespective of the strength, 
Fermi surface of the degenerate Fermi gas is unstable with respect to the formation of two particle 
bound state, a {\it Cooper pair}, and the system turns into a coherent mixture of such bound states 
described by the BCS wave function.    Similar situaton may arise in the degenerate quark plasma 
if there is attractive quark-quark interaction at the Fermi surface; this happened to be the case for 
anti-triplet color channel of quark pair and this problem has been extensively studied 
in recent years.\cite{RW01}

Fig. 1 summarizes the present theoretical expectation of the phase diagram of hot and dense matter as
plotted in the plane of the temperature $T$ and the baryon chemical potential $\mu$, the measure of 
the asymmetry of the baryon-antibaryon abundance in the system.

\section{Nuclear collisions at ultrarelativistic energies}

\begin{table}[htdp]
\caption{Relativistic heavy-ion accelerators}
\begin{center}
\begin{tabular}{lrrrr}
\hline
&$E_{Lab}$ [GeV] &$E_{cm} [GeV] $
&$\Delta y$ & $\gamma_{cm}$
\\
\hline
AGS (BNL) & 30$Z/A$  & 6$Z/A$ & 3.5 & 7 \\
SPS (CERN) & 400$Z/A$ & 20$Z/A$ &6 & 10 \\
RHIC (BNL) & -- & 250$Z/A$ & 11 & 100 \\
LHC (CERN) & -- & 8000$Z/A$ &  17 & 2500 \\
\hline
\end{tabular}
\end{center}
\label{accelerator}
\end{table}%

The only method at our disposal, although very crude and not ideal, to create and study hot dense matter 
in laboratory experiments is to collide two nuclei at extreme relativistic energies.  
Such high energy nuclear collisions have been observed since more than a half century ago in 
high energy cosmic ray events as a special cases of multiparticle production phenomena which 
have been described in terms of phenomenological thermodynamic or hydrodynamic models incorporating 
space-time picture as imposed by the special relativity.\cite{Fer50} 
The interests in high energy nucleus-nucleus collisions revived in 80's from the prospect of creating 
and studying a quark-gluon plasma and this led to the initiation of the experimental programs at Brookhaven 
and CERN in mid 80's using existing hadron accelerators (AGS at Brookhaven and SPS at CERN), 
while a dedicated relativistic heavy-ion collider (RHIC) was built at Brookhaven.  The energy per 
nucleon of ion beams accerelated by these machines are shown in table 1. 
In the table, the center of mass collision energies per nucleon $E_{cm} = \sqrt{m_N E_{Lab}/2}$
of AGS and SPS fixed target experiments are given together with the rapidity difference of target and
projectile nucleons ($\Delta y = y_{p} - y_{t}$).  The table also contains the parameters of 
LHC which is now under construction at CERN.

At untrarelativistic energies, a head-on collision of two nuclei mass number $A$ may be view as 
a collision of two highly contracted nuclei of a disk shape.  
The Lorentz contraction factor in the center of mass frame, $\gamma_{cm} = E_{cm}/m$,
is $\gamma_{cm} \simeq 100$  at the RHIC energy and the longitudinal thickness of the nuclear
disk becomes $2R/\gamma_{cm} \simeq 0.015$fm even for heavy nuclei, which is smaller than 
the hadron size.   Under such conditions, it may be more natural to view the collision process as 
a collision of two inter-penetrating beams of partons, the constituents of nucleons,  
instead of collisions of individual nucleon as a whole.  
After the collision, two "wounded nuclei" consisting of these "primary" partons would continue to fly 
along essentially the same paths as the free particles, but the space-time region sandwitched by 
them will be excited and filled up with excitations of quarks and gluons.\cite{Bj76}
The net baryon number will be carried away with the primary partons and hence we may expect 
the formation of quark-gluon plasma with vanishing baryon chemical potential as in the very
early universe.   Matter created in between two receeding nuclei will expand and cool down, and 
it will eventually disassemble into ordinary hadrons, and a few leptons, filling the mid-rapidity
region of particle distribution in the rapidity space.  A space-time view of matter evolution seen 
in the center of mass frame is depicted in Fig. 2.

\begin{figure}[t]
  \epsfxsize=6cm  
  \centerline{\epsfbox{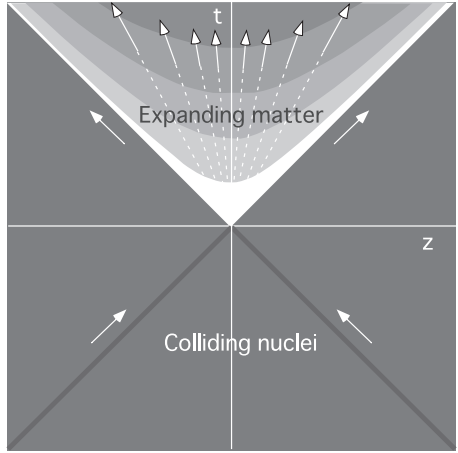}}
  \caption{Space-time view of a nuclear collision}
   \label{fig2}
\end{figure}

This dense gas of (secondary) partons may, or may not, quickly achieve a local thermodynamic 
equilibrium by their mutual interaction.  If it does, as first assumed by Landau, then the subsequent 
evolution of the system will be described entirely by the conservation laws 
provided that the equation of state of equilibrium matter is known; then an adiabatic 
hydrodynamic expansion of the system will follow.  

The first stage of the adiabatic expansion is of one-dimensional character since the expansion is
predominantly in the longitudinal direction and the cooling of the system will take place as the 
conserved entropy of the system will be diluted by being spread over an expanding volume 
as the matter expands.  
This adiabatic cooling is determined by solving the hydrodynamic equations, but if we adopt 
the Lorentz boost invariant expansion, along with Bjorken\cite{Bj83}, which ensures the Ansatz 
$v_z = z/t$ for the longitudinal flow velocity and $\tau = \sqrt{ t^2 - z^2 }$ for the proper time, 
then we find that the entropy density decreases inversely proportional to $\tau$.   
This results in $T \propto \tau^{1/3}$ if we use the ultrarelativistic ideal gas relation between 
the entropy density $s$ and the temperature $T$: $s = a T^3$ where $a = 2 N_{eff.} \pi^2/ 45$ and 
the effective degrees of freedom is given by $N_{eff} = 16 + 21 n_f/2$ for an ideal gas of color 
octet gluons and massless $n_f$ flavor quarks.  During this stage the transverse expansion of the
system sets in from the outer edge of the system where the pressure gradients generate a 
transverse acceleration of matter.   The inward edge of the transverse rarefaction wave propagate
through the longitudinally expanding matter with the velocity of sound and an element of matter 
will be set in transverse motion after the wave passes by as illustrated in Fig. 2.   

As the rarefaction wave reach the center of the matter, the whole matter will be set in full
three dimensional expansion and it will disintegrate into a free stream of hadrons quickly. 
This time scale is given in terms of nuclear radius $R$ and the sound velocity $c_s$ as 
$\tau_{exp} = R/c_s \simeq (1 - 2 ) R$ fm/c.  
The lifetime of the plasma is determined by this expansion time scale, although 
detail of its evolution is influenced by other factors like the kinetic properties of the plasma 
and its hadronization mechanism which are still unknown and remain as the challenging problems 
of theoretical research.

\begin{figure}[t]
  \epsfxsize=6cm  
  \centerline{\epsfbox{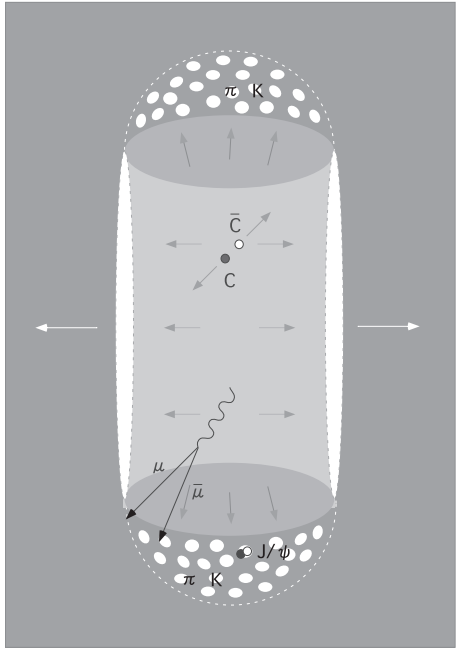}}
  \caption{A snapshot of expanding matter created by nuclear collision}
   \label{fig3}
\end{figure}

\section{Probes of the quark-gluon plasma}
As we have discussed in the previous section, even if a quark-gluon plasma is produced 
in high energy nuclear collision it will cool down very rapidly and will disassemble 
into thousands of ordinary hadrons on a very short time scale of $10^{-21} - 10^{-20} $ seconds. 
This makes it extremely difficult to identify the signal of the quark-gluon plasma formation.
Many ideas have been proposed and some of the important ones will be discussed now 
in the light of experimental data taken after the proposal.

{\bf Flavor composition}: 
In a quark-gluon plasma, quarks and anti-quarks are populated according to the statistical rule.  
In equilibrium, it is determined only by the mass of the excitations and there is a good reason 
to assume that three light flavor quarks (up, down, strange quarks) are almost equally abundant 
in the plasma with zero net baryon number.  
The time scale for the approach to flavor equilibrium was computed by perturbative QCD and was 
found to be shorter than the lifetime of the system\cite{RM82}.  It was soon realized\cite{KM86}, 
however, that this initial symmetry in flavor abundance may not be reflected directly in the final 
hadron abundance as an enhancement of the production of strange hadrons 
($K$, $\Lambda$, $\bar \Lambda$, ...); contrary to this naive expectation, if a quark-gluon plasma 
is formed and adiabatically evolved into hadron gas, its entropy content should be preserved and 
most likely this will lead to a dilution of the $K/\pi$ ratio because pions are most easily produced 
to compensate the entropy of gluons.  The final relative abundance of various hadrons may only 
reflect the freeze-out stage of the matter.

Enhancement of strange particle production was observed both in AGS\cite{E802} and
SPS\cite{WA97} experiments.  Most part of these data are fitted very well to a simple 
statistical abundance of ideal gas with two fitting parameters $T$ and $\mu$.\cite{BHS99}
It remains an interesting question whether an anomalous yield of some very exotic hadrons such 
as $\Omega^-$ which consists of three strange quarks and therefore is very difficult to be produced 
in hadron-hadron collision may be enhanced anomalously in a nuclear collision was indeed observed 
at SPS experiments.

{\bf Leptons and photons}:
Leptons and photons are called {\it penetrating probes} since they are free from strong final state
interaction and thus expected to carry information of the interior of the matter produced. 
\cite{Fei76} In 
contrast, hadrons are created only at the surface (hyperspace in 3 + 1 dimensional Minkowski 
space) of the plasma and may suffer more final state interaction before they freeze out into free 
streaming particles. In particular, dileptons (a pair of lepton and its anti-particle) couple to local
thermal fluctuation of electromagnetic current in the hot matter and their invariant mass spectrum
may reveal the nature of the matter from which they are emitted.  The high mass tail of the 
dilepton invariant mass spectrum $M > 3 \rm{~GeV}$ is known to be dominated by the "Drell-Yan" 
pairs which are produced at the earliest stage of the nuclear collision by annihilation of primary 
partons (quark and anti-quark) into a virtual photon.   Dilepton created in the interior of 
the quark-gluon plasma by the annihilation of thermally excited quark and anti-quark pair are
expected to dominate in the intermediate mass range $2 \rm{~GeV} < M < 3 \rm{~GeV}$,\cite{MT85} 
while many of the low mass dileptons are produced by the electromagnetic decay of hadrons
and hence interesting as a probe of properties of individuals hadron in the dense medium
as expected from some theoretical models.\cite{HL92} 

The data taken at SPS indeed shows some enhancement and change in the shape of low 
mass dilepton spectrum in $e^+e^-$ channel\cite{CERES}, however the effect is seen
in light-ion induced reactions and no more prominent in heavy-ion induced 
reactions.  Also it can be understood in terms of many-body correlations in the 
medium\cite{WR98} without invoking the change of hadronic properties.  

{\bf Quarkonium}: 
A pair of heavy quark and its anti-particle ($c \bar c$, $b \bar b$) are occasionally 
produced at the initial stage of collision by primary parton interaction and some of 
them can evolve into a bound state called quarkonium. 
Hadronic production of vector quarkonium can be observed by the high mass resonances
in the dilepton invariant mass spectrum.
In particular, $J/\psi$ ($^3S_1$ $c \bar c$ state) forms a prominent peak 
at $M = 3.1 $GeV which stands out of Drell-Yan continuum.  
If a pair is produced in the event which also results in formation of a quark-gluon 
plasma, then the subsequent evolution of the pair to bound state will be prohibited 
by the plasma screening of the mutual interaction of the pair.  Since the continuum 
Drell-Yan pair will not be much affected by the plasma formation, this should result
in a strong suppression of the peak/continuum ratio in the dilepton spectrum:
$J/\psi$ suppression.\cite{MS86}
So we proposed that the signature is an absence of a signature. 
Mass shift of the $J/\psi$ peak was also proposed\cite{HHKM86} as a signal of 
precursory effect of deconfinement, but unfortunately the long time scale for the decay 
of $J/\psi$ prohibits to observe this effect. 

Suppression of the $J/\psi$ peak relative to the Drell-Yan continuum was observed in the early
experiments with light ions at SPS.\cite{NA38}  
However, it was soon found\cite{GH92} that the observed suppression
can be  interpreted in terms of a "nuclear absorption" model which parameterizes the 
collisional loss of charmonium on the way out of nuclei, the effect already seen in 
$pA$ collisions.\cite{pA}  A new surprise came with the data from the lead beam experiments 
which exhibit anomalous suppression in the events at small impact parameters\cite{NA50}
which cannot be fitted by extrapolation by simple nuclear absorption model
and requires additional exotic mechanism such as plasma suppression.\cite{BO96,KLNS97,HS00}
This conclusion was challenged by the claim\cite{GG99} that the observed yields of $J/\psi$ 
as compared to other lighter hadrons are very close to the statistical equilibrium value which 
would mean that most of $J/\psi$'s are produced by the recombination of $c \bar c$ from a
thermal bath.  
Although thermal origin of $J/\psi$ at SPS energies is not consistent with other data\cite{BS00}, 
it was pointed out that there would be an enhancement at higher collider energies.\cite{BS00,TSR01} 
A preliminary RHIC data seem to exclude this possibility.\cite{Phenix}

{\bf Jets}: Hard scatterings of primary partons generates a pair of energetic partons carrying
large transverse momenta which will fragment into back-to-back hadron jets in the free space.  
Such jets have been well identified in $p \bar p$ collisions at Tevatron.\cite{CDF} 
In nucleus-nucleus collision each member of the pair scattered in deep interior of the collision 
volume will travel through the matter on its way out before fragmenting into hadrons and 
thus will change its energy-momentum by interaction with the medium.
The distances that two members of the pair travel depend on the location where the primary parton 
scattering took place in the collision volume and therefore they are not the same in most cases. 
This may lead to an imbalance in the two jets or even extinction of one of the jets or 
both.\cite{Bj82}
Significance of this effect depends crucially on the parton energy loss in the dense matter which 
depends on the nature of matter the high energy partons traverse.   
It was pointed out that the gluon radiation is a dominant mechanism of the parton energy loss 
in the quark-gluon plasma\cite{GP90} 
and it leads to thermalization of mini-jets components which otherwise dominate the high momentum
tail of the hadron spectrum.\cite{WG92}

The inclusive hadron spectrum observed in SPS experiments shows enhancement at high momentum and
this has been interpreted as a multiple scattering effect ("Cronin effect" in $pA$ collision)  or 
hydrodynamic flow effect.
The data taken at RHIC however have qualitatively different feature from these SPS data, showing 
systematic suppression of the high momentum tail at $E > 5$ GeV which is suggestive of a manifestation of 
jet quenching.\cite{Hirano}

\section{Outlook}
We have learned so many things from the fixed target experiments at AGS and SPS, and some
of the data strongly suggest a picture that a quark-gluon plasma is formed at SPS at least for a 
short while.
It is, however, still very important to confirm these results by on-going RHIC experiments and future
experiments at LHC more systematically and to look for new effects which become manifest only
at high energies.
As we increase the energy of collisions, hard scattering of primary partons plays more and more 
important role.  
It has been anticipated\cite{Esk97} that "mini-jet" components will eventually dominate 
the energy deposition which at lower energies is mainly due to the "soft" interactions 
as modelled by string formation and decay. 
It has also been emphasized that the rescatterings of these hard partons in dense partonic medium 
(quark-gluon plasma) are crucial to assess the utility of various probes of the quark-gluon 
plasma.\cite{Wan97}
Increase of the collision energy is generally favorable to create a denser (hotter) matter with a 
longer lifetime. 

We should also keep in mind that the physical conditions achieved by these new collider 
experiments are very different from those of fixed target experiments, and some of the assumptions
implicitly used for the analysis of SPS data may not be applicable for interpretation of new data. 
For example, we have recently pointed out that the nuclear absorption of charmonium will change 
qualitatively at high energies due to quantum coherence in multiple scatterings.\cite{FM02,Fujii} 
Increase of charm production rate may become a serious problem to use the charmonium suppression 
as a probe of the quark-gluon plasma at LHC.   
This needs to be checked experimentally, but if this turns out to be the case we may resort to 
bottomonium ($b \bar b$ bound states like $\Upsilon$) as a new probe of the very high temperature 
plasma. 

In closing, I like to quote a very famous phrase from Chinese translation of a sutra 
of Mahayana Buddhism: 

\begin{figure}[h]
  \epsfxsize=4cm  
  \centerline{\epsfbox{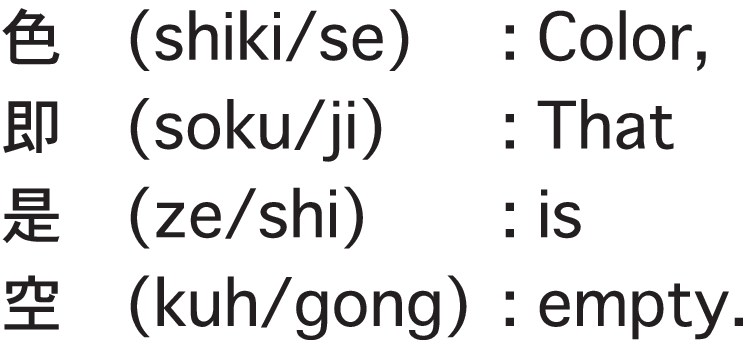}}
   \label{fig4}
\end{figure}

\vskip -5pt

\noindent A word-to-word translation of the phrase is indicated on the
right of each word together with Japanese/Chinese pronounciation.  
It would imply:  "Thing which has color does not exist."  
One may be surprized at finding that the ancient Buddhist teaching contains such a modern 
statement on the quark confinement! \cite{opposite}
The true meaning of this phrase is said to be that "Everything around us, 
which has color so that we can see, does not stay unchanged eternally.  
Things exist only in a state of flux. "  
This statement is considered as an essence of the Buddhist view of the world which
may look somewhat pessimistic from the point of view of scientific endeavor.  
In science, we are always looking for knowledge of lasting significance. 
I may still conclude this talk happily in accord with this phrase as follows:
Our Universe was created in the Big Bang and its matter content has evolved from the primeval 
plasma to the present state of matter, as we see it. 
The quark-gluon plasma may still be formed in a state of flux in nuclear collision and
we may have already seen a flash of it!

\section*{Acknowledgements}
The author thanks the organizer of the workshop, Tetsuo Hatsuda, for his kind invitation and my
colleague in Komaba, Hirotsugu Fujii, for collaboration
and helpful conversations on the most recent 
results from RHIC experiments.


\begin{thebibliography}{99}
\bibitem{opposite} The opposite statement: 
\begin{figure}[h]
\vskip -0.82cm
  \epsfxsize=4cm  
  \centerline{\epsfbox{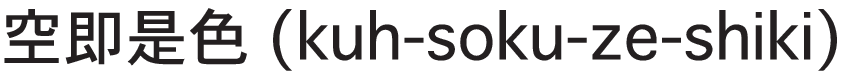}}
   \vskip -0.45cm
\end{figure}
\hskip 4.5cm which is also written in the sutra can find another
interesting modern interpretation in QCD physics: 
"the vacuum is a (non-trivial) state of color (fluctuation)!".
\end{thebibliography}
\end{document}